\documentclass[aps,pra, twocolumn, superscriptaddress, showpacs, reprint]{revtex4-1}

\usepackage{amsmath,amssymb}
\usepackage{xcolor}
\usepackage{graphicx}
\usepackage{braket}
\usepackage[utf8]{inputenc}
\usepackage{bm}

\newcommand\Id{\openone}

\begin{document}
\date{\today}

\title{Measuring Topological Invariants in Disordered Discrete Time Quantum Walks}
\author{Sonja Barkhofen}
\affiliation{Applied Physics, University of Paderborn, Warburger Strasse 100, 33098 Paderborn, Germany}
\author{Thomas Nitsche}
\affiliation{Applied Physics, University of Paderborn, Warburger Strasse 100, 33098 Paderborn, Germany}
\author{Fabian Elster}
\affiliation{Applied Physics, University of Paderborn, Warburger Strasse 100, 33098 Paderborn, Germany}
\author{Lennart Lorz}
\affiliation{Applied Physics, University of Paderborn, Warburger Strasse 100, 33098 Paderborn, Germany}
\author{Aur\'el G\'abris}
\affiliation{Department of Physics, Faculty of Nuclear Sciences and
  Physical Engineering, Czech Technical University in Prague, B{\v r}ehov\'a
  7, 115 19 Praha 1--Star\'e M{\v e}sto, Czech Republic}
\affiliation{Department of Theoretical Physics, University of Szeged, Tisza Lajos k\"or\'ut 84, H-6720 Szeged, Hungary}
\author{Igor Jex}
\affiliation{Department of Physics, Faculty of Nuclear Sciences and
  Physical Engineering, Czech Technical University in Prague, B{\v r}ehov\'a
  7, 115 19 Praha 1--Star\'e M{\v e}sto, Czech Republic}
\author{Christine Silberhorn}
\affiliation{Applied Physics, University of Paderborn, Warburger Strasse 100, 33098 Paderborn, Germany}

\begin{abstract}
Quantum walks constitute a versatile platform for simulating transport phenomena on discrete graphs including topological material properties while providing a high control over the relevant parameters at the same time.
To experimentally access and directly measure the topological invariants of quantum walks we implement the scattering scheme proposed by Tarasinski et al.\ [Phys.\ Rev.\ A 89, 042327 (2014)] in a photonic time multiplexed quantum walk experiment.
The tunable coin operation provides opportunity to reach distinct topological phases, and accordingly to observe the corresponding topological phase transitions.
The ability to read-out the position and the coin state distribution, complemented by explicit interferometric sign measurements, allowed the reconstruction of the scattered reflection amplitudes and thus the computation of the associated bulk topological invariants.
As predicted we also find localised states at the edges between two bulks belonging to different topological phases.
In order to analyse the impact of disorder we have measured invariants of two different types of disordered samples in large ensemble measurements, demonstrating their constancy in one disorder regime and a continuous transition with increasing disorder strength for the second disorder sample.

\end{abstract}

\pacs{
03.67.Ac,
42.50.-p,
03.65.Vf}

\maketitle

\section{Introduction}
Quantum walks \cite{aharonov_quantum_1993, meyer_absence_1996, farhi_quantum_1998} have gained recognition as a potential platform for quantum algorithms \cite{childs_example_2002, shenvi_quantum_2003, ambainis_coins_2005, ambainis_quantum_2007, childs_universal_2013} and quantum simulators \cite{strauch_relativistic_2006, witthaut_quantum_2010, engel_evidence_2007, mohseni_environment-assisted_2008, lee_quantum_2015}.
In the last years, they have been realized experimentally in various systems, demonstrating different effects known from condensed matter physics such as dynamical localization \cite{schreiber_decoherence_2011, schreiber_photons_2010, genske_electric_2013, crespi_anderson_2013}, percolation \cite{elster_quantum_2015}, interacting systems \cite{schreiber_2d_2012, preiss_strongly_2015}, and more recently, topological insulators \cite{rechtsman_photonic_2013, kitagawa_observation_2012, poli_selective_2015, zeuner_observation_2015, cardano_dynamical_2015}.
Topological insulators have received substantial attention due the emergence of protected bound states localized on the edge of such samples (see \cite{hasan_colloquium_2010} for a recent review).
Such robust modes constitute an attractive basis for real-world applications in quantum technology, and quantum information related fields \cite{hsieh_tunable_2009,moulieras_entanglement_2013}.
Recent findings of Kitagawa et\,al.~\cite{kitagawa_exploring_2010, kitagawa_topological_2012} regarding discrete time quantum walks (DTQW) have been carried further to show that in general Floquet systems can exhibit topological features \cite{jiang_majorana_2011, rudner_anomalous_2013, fulga_scattering_2016}.
In the work by Cardano et~al., which appeared during the preparation of our manuscript, they measured the mean chiral displacement, a quantity which converges to the Zak phase, in a photonic DTQW implemented by the orbital angular momentum of a light beam over 7 steps and extracted both topological invariants, also in the presence of disorder in ensemble averages over 10 different patterns~\cite{cardano_detection_2017}.

The theoretical study of topological effects in the context of the DTQW turned out to be a fruitful subject marked by works of Asb\'oth \cite{asboth_symmetries_2012, tarasinski_scattering_2014,asboth_bulk-boundary_2013} and Werner \cite{cedzich_bulk-edge_2016, cedzich_topological_2016}, and is still open for debates.
A method to determine bulk topological invariants based on scattering has been proposed \cite{fulga_scattering_2011} and realized \cite{hu_measurement_2015} recently.
Its subsequent adoption to the DTQW with certain modifications \cite{tarasinski_scattering_2014} yields a robust and reliable tool for the verification of the existence of topological phases in experimentally realized synthetic quantum systems.
Such realizations are expected to constitute the first step towards applications in quantum technology.
While localised states in this context have already been observed experimentally in DTQW systems ~\cite{kitagawa_observation_2012}, a thorough experimental proof of manifestation of topological phases and their behaviour in disordered systems had been missing.

In this article, we present the results from three series of measurements conducted using an optical time-multiplexed feedback loop following the scattering scheme proposed in Ref.~\cite{tarasinski_scattering_2014} to experimentally determine the topological invariants of one dimensional split-step quantum walks.
In the first series we determine the topological invariants of disorder-free systems comprising a lead channel connected to a bulk sample, which we tune from one phase to another in order to observe topological phase transitions.
In the second series of measurements we make use of the capability of our apparatus to introduce binary disorder by randomly replacing the original coin operator of the bulk sample by another one for each single position.
The strength of the randomness is determined by the probability $p$ for choosing the new coin instead of the original one.
This also requires an averaging over 50 coin configurations for each $p$.
Our study of disordered quantum walks covers two complementary cases, one when the original and the new coin operator belong to the same topological phase, and a second when they are picked from different ones.
In the first case study we observe constant topological invariants independent of the degree of disorder, while in the second case study we observe a clear dependence of the topological invariants on the disorder strength.
Third, we experimentally demonstrate the correspondence between the topological invariants from the described scattering systems and the existence of localised edge states on the boundary of two bulk samples.

\section{Theoretical Background}
In analogy to its classical counterpart---the random walk---the dynamics of the usual DTQW is given by the alternating application of a coin toss $\hat{C}$ and a conditional shift $\hat{S}$ in space.
In the quantum case, the walker can be in a superposition state of positions $x$ and coin states $c$ given by 
\begin{equation}
| \Psi \rangle_t =  \sum_x \sum_c a_{x,c}(t)| x \rangle \otimes | c\rangle
\end{equation}
with the time-dependent amplitudes $a_{x,c}(t) \in \mathbb{C}$.
Then, the dynamics can be described as
\begin{equation}
| \Psi \rangle_{t+1} = \hat{U}| \Psi \rangle_t= \hat{S}\hat{C}| \Psi \rangle_t,
\end{equation}
with $\hat{C}=\Id_x \otimes \hat{R}(\theta) = \Id_x \otimes e^{-i\hat{\sigma}_x \theta}$ being Pauli rotations in the coin space expressed in the horizontal and vertical basis states $|H\rangle = (1~0)^T$ and $|V\rangle = (0~1)^T$ and the shift operator 
\begin{equation}
\hat{S} = \sum_x \ket{x + 1}\!\bra{x}\otimes \ket{H}\!\bra{H} + \sum_x \ket{x - 1}\!\bra{x}\otimes \ket{V}\!\bra{V}~.
\end{equation}
Note, that the choice of gauge of the coin operator is motivated by the experimental implementation and differs from that used in the theory literature e.g.\ in Ref.~\cite{kitagawa_exploring_2010}.
In the rest we follow the literature and adopt the split-step protocol, where the unitary operator of a single time step is constructed as
\begin{equation}
  \hat{U} =  \hat{S}_- \hat{C}_2 \hat{S}_+ \hat{C}_1,
	\label{eq:splitstepU}
\end{equation}
with two coin operators $\hat{C}_{1,2}$, and the asymmetric shift operators 
\begin{eqnarray}
\hat{S}_{+} &=& \sum_x \ket{x + 1}\!\bra{x}\otimes \ket{H}\!\bra{H} + \Id_x \otimes \ket{V}\!\bra{V} \\
\hat{S}_{-} &=& \Id_x \otimes \ket{H}\!\bra{H} + \sum_x \ket{x - 1}\!\bra{x} \otimes \ket{V}\!\bra{V}~.
\end{eqnarray}
Note, that the split-step scheme is equivalent to two time steps of the original DTQW given by $\tilde{U}=\hat{S}\hat{C}_2\hat{S}\hat{C}_1$, i.e.\ two spatially alternating coin operators and suitable relabelling of the positions (see appendix~\ref{App:splitstepscheme}).
We make use of this equivalence for the experimental realisation of the protocol.

\begin{figure}
 \centering
 \includegraphics[width=0.8\columnwidth]{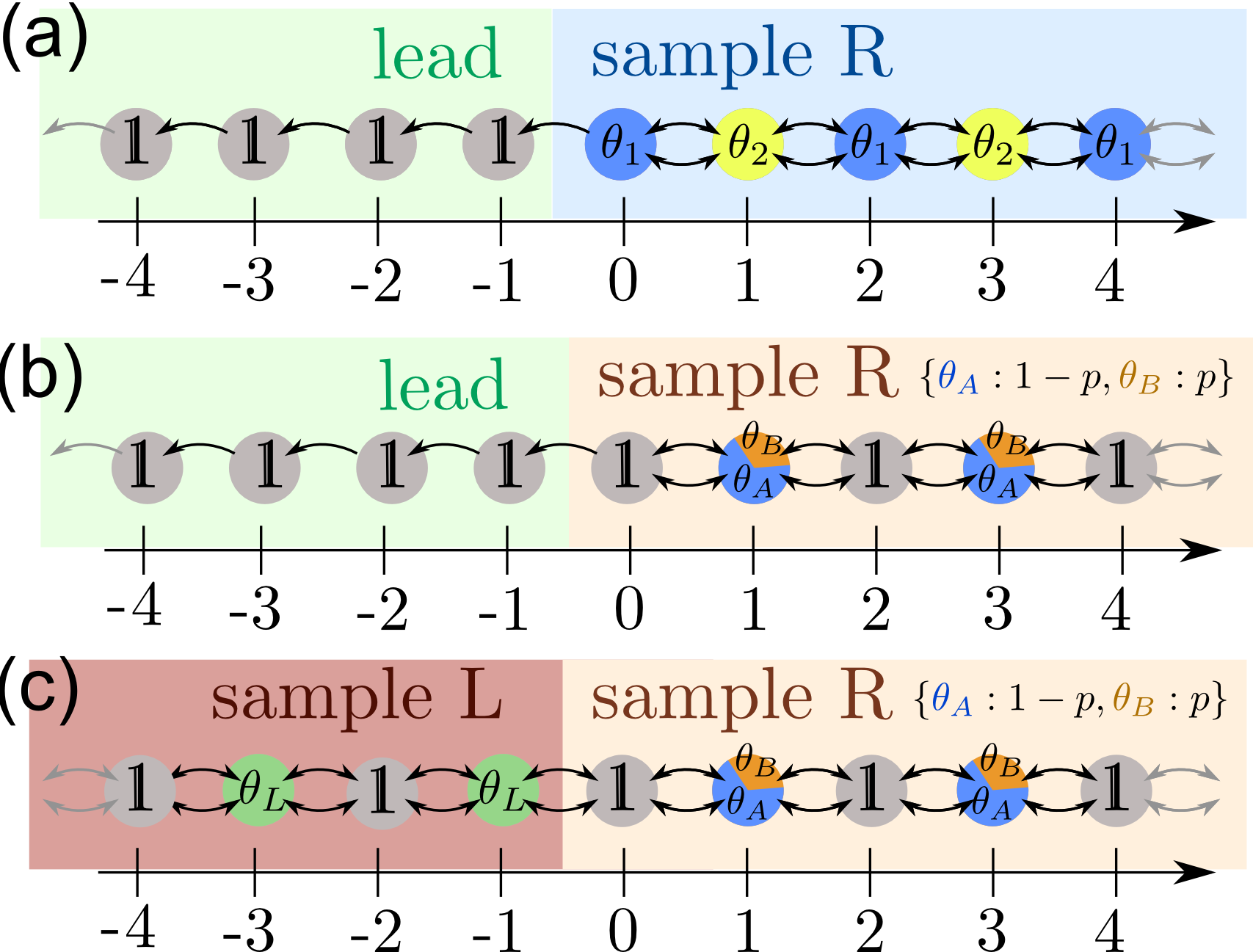}  
 \caption{\label{fig:leadBulk}   
(a) Lead--sample scattering system: sample to the right with alternating coins $\hat{C}_1 = \openone_x \otimes \hat{R}(\theta_1)$ and $\hat{C}_2 = \openone_x \otimes \hat{R}(\theta_2)$ connected to a lead comprising only identity coins. 
(b) now sample with $\hat{C}_1=\openone$ being constant, and $\hat{C}_2$ at each site chosen from $\hat{R}(\theta_{\mathrm{A}})$ and $\hat{R}(\theta_{\mathrm{B}})$, with respective probabilities $1-p$ and $p$, yielding disordered systems for $p\neq 0,1$. 
(c) Interface between two samples with $\hat{C}_1=\openone$. The left sample consisting of $\hat{C}_2=\openone_x \otimes \hat{R}(\theta_L)$ is connected to the same disordered samples as in (b).}
\end{figure}
\begin{figure}
 \centering
 \includegraphics[width=\columnwidth]{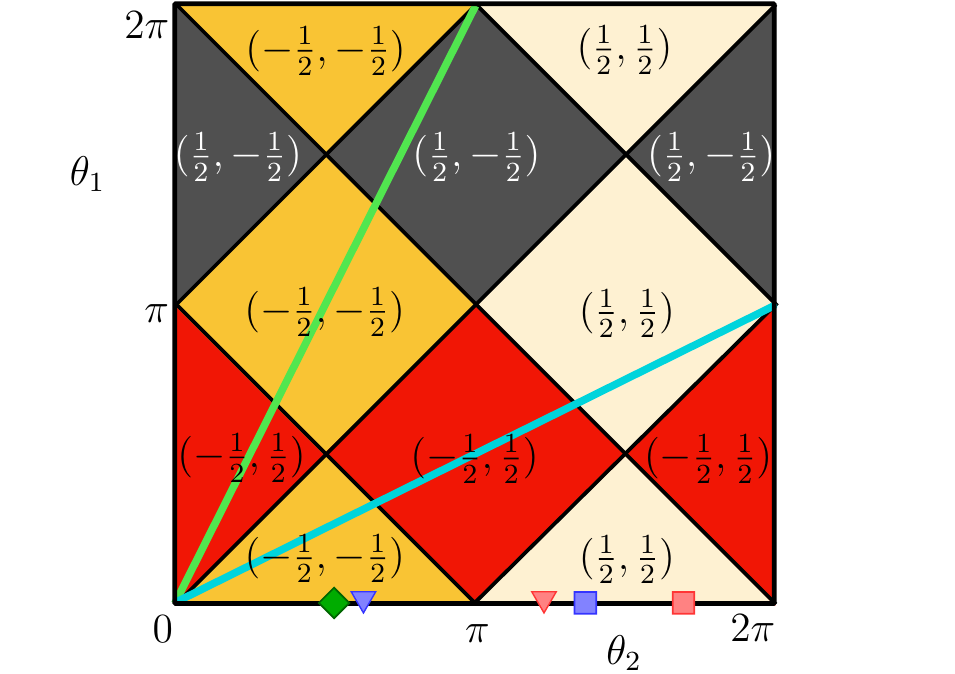}    
 \caption{\label{fig:parameterPlot} Topological phase diagram indicating the bulk topological invariants ($Q_0$, $Q_{\pi}$) as functions of the coin angles $\theta_1$ and $\theta_2$ of a split-step quantum walk. The colored lines and symbols indicate the parameters of the experimentally realised systems.}
\end{figure}

To directly measure the topological invariants we follow Ref.~\cite{tarasinski_scattering_2014} and study a scattering arrangement comprising a half-infinite lead with trivial dynamics ($\hat{U}_{\mathrm{lead}}=\hat{S}_{+}\hat{S}_{-}$) to the left, and a half-infinite sample (a generic split-step quantum walk with non-trivial coin operations) to the right as depicted on Fig.~\ref{fig:leadBulk}(a).
The generic split-step walk constitutes the sample of which the topological invariant is to be determined.
It is characterised by the position dependent coin operator $\hat{C}({\bm\theta}) = \sum_x \ket{x}\!\bra{x} \otimes \hat{R}(\theta_x)$ with the formal vector ${\bm \theta}= (\ldots, \theta_{x=0}, \theta_{x=1},\theta_{x=2}, \ldots)$ describing the spatial arrangement of the coin angles.

The central object in our study is the quasi-energy dependent scattering matrix, which describes how an incoming plane wave, or strictly speaking its quantum-walk equivalent, at the respective energy travelling through the lead is scattered off the sample and thus partially reflected and transmitted into outgoing modes.
Consequently, the scattering matrix consists of reflection and transmission matrix elements, each one dependent on the energy. 
The DTQW, being a Floquet system in contrast to Hamiltonian systems usually describing topological insulators, e.g.~the SSH-model \cite{su_solitons_1979}, has relevant quasi-energy gaps at $\varepsilon=0$ and $\varepsilon=\pi$ in which topologically protected states may appear \cite{jiang_majorana_2011}.
The transmission amplitudes at the two quasi-energies are exponentially small for growing sample size, such that the reflection matrices $r(0)$ and $r(\pi)$ become unitary.
If a scattering system possesses time-reversal, particle-hole or chiral symmetry, as used in the standard band theory of topological insulators, the action of the present symmetries, in the following denoted by (anti-)unitary operators $\tau$, $\eta$ and $\gamma$, respectively, on the in- and out-going waves in the lead is mirrored in the reflection properties of the sample.
For example time-reversal symmetry will reverse the action of the time-evolution operator $\hat{U}$, and thus map incoming waves to outgoing
waves and vice versa.
The main findings of \cite{tarasinski_scattering_2014} are that the topological invariants are obtained as simple functions, e.g. trace, determinant or Pfaffian, of the scattering matrix, strictly speaking its reflection blocks, at these quasi-energies.
Thus the invariants come in pairs, here denoted by $(Q_0,Q_{\pi})$, and are unique up to a global multiplicative factor \cite{asboth_bulk-boundary_2013, rudner_anomalous_2013, fulga_scattering_2016}. 
The explicit relation of the energy dependent reflection element of the scattering matrix $r(\varepsilon)$ to the invariants, as well as their possible values are determined by the symmetry class of the particular sample.
The split-step walk studied in this article belongs to the BDI symmetry class \cite{asboth_bulk-boundary_2013}, which means that all three symmetries are present, square to $+1$ and interact with the time-evolution operator $\hat{U}$ of the walk, defined in eq.~(\ref{eq:splitstepU}), as $\tau \hat{U}\tau^* = \hat{U}^*$ (time-reversal), $\eta \hat{U}\eta^* = \hat{U}$ (particle-hole) and $\gamma \hat{U}\gamma^* = \hat{U}^*$ (chiral symmetry).
Due to these symmetries the reflection blocks of the scattering matrix are Hermitian (and unitary) and thus their eigenvalues are given by $\pm1$.
This leads to the explicit formula for the topological invariants related to the trace of $r(\varepsilon)$:
\begin{equation}
(Q_0,Q_{\pi}) = \frac12 (\mathrm{Tr}~r(0), \mathrm{Tr}~r(\pi))~.
\end{equation}
For a 1d walk, as studied here, the reflection blocks of the scattering matrix are also one-dimensional and the formula for the invariants simplifies to 
\begin{equation}
(Q_0,Q_{\pi}) = \frac12 (r(0), r(\pi))
\label{eq:Q0Qpi}
\end{equation} 
and yields the topological phase diagram shown on Fig.~\ref{fig:parameterPlot}.

We shall denote by $r_j$ the reflection amplitude at position $x=-1$, i.e. the border between lead and sample, after $j$-times application of $\hat{U}$ on an initially localized state $\ket{H}$ at $x=0$, i.~e.
\begin{equation}
  r_j = \Bra{-1,V} \hat{U}^{j} \Ket{ 0, H },
  \label{eq:refl_amplitudes}
\end{equation}
following the labelling convention depicted on Fig.~\ref{fig:leadBulk}(a).
Following Ref.~\cite{tarasinski_scattering_2014}, the energy dependent reflection matrix element can be expressed as the discrete Fourier transform of the reflection amplitudes given by
\begin{equation}
  r(\varepsilon) = \sum_{j=1}^{t \rightarrow \infty} e^{ij\varepsilon} r_j.
  \label{eq:refl_matrix}
\end{equation}
which simplifies to the sum and alternating sum for the relevant energies in eq.~(\ref{eq:Q0Qpi}), i.e. $\epsilon = 0$ and $\epsilon = \pi$, respectively.
Since the $r_j$ are in general measurable quantities we can thus easily infer the topological invariants by making use of eqs.~(\ref{eq:Q0Qpi}) and (\ref{eq:refl_matrix}).
The extent of control in presented experimental realization of DTQW make up to a well suited platform for studying topological invariants in such a scattering setting.
In particular, the dispersion-free nature of the lead implementation allow for a single shot generation of all required reflection amplitudes using an initially localized state.
Nevertheless, the experimental reconstruction of the amplitudes from measured intensities (count rates) still requires several repetitions to ensure statistical significance.
Moreover, in present implementations of DTQW the number of steps is limited, and in finite measurement time the reflection matrix can be recovered only approximately.
It is the speed of convergence of the sum in eq.~(\ref{eq:refl_matrix}) with finite $t$ which determines the accuracy of the experiment employing finite steps.

\section{Experimental Realization}
We implement a photonic quantum walk applying a fibre loop architecture based on the time-multiplexing technique providing resource efficiency, high homogeneity and long-lasting stability against uncontrolled dephasing, see \cite{schreiber_photons_2010, elster_quantum_2015}.
A weak coherent laser pulse plays the role of the walker and its polarization represents the coin state.
In combination with standard static linear elements, fast electro-optic modulators (EOM) can be used to modify the polarisation of each pulse individually \cite{schreiber_decoherence_2011, elster_quantum_2015}, providing the basis for realising position dependent coin operations $\hat{C}({\bm\theta})$.
The step operation $\hat{S}$ (or $\hat{S}_{\pm}$ \footnote{In this case the choice of convention is a matter of taste}) is performed by routing the two polarisation components through fibres of different lengths to introduce a well-defined time delay between them.
Consequently, each position in each roundtrip is uniquely represented by discrete time bins, i.e.\ the position information is mapped into the time domain.

\begin{figure}
\centering
 \includegraphics[width=0.95\columnwidth]{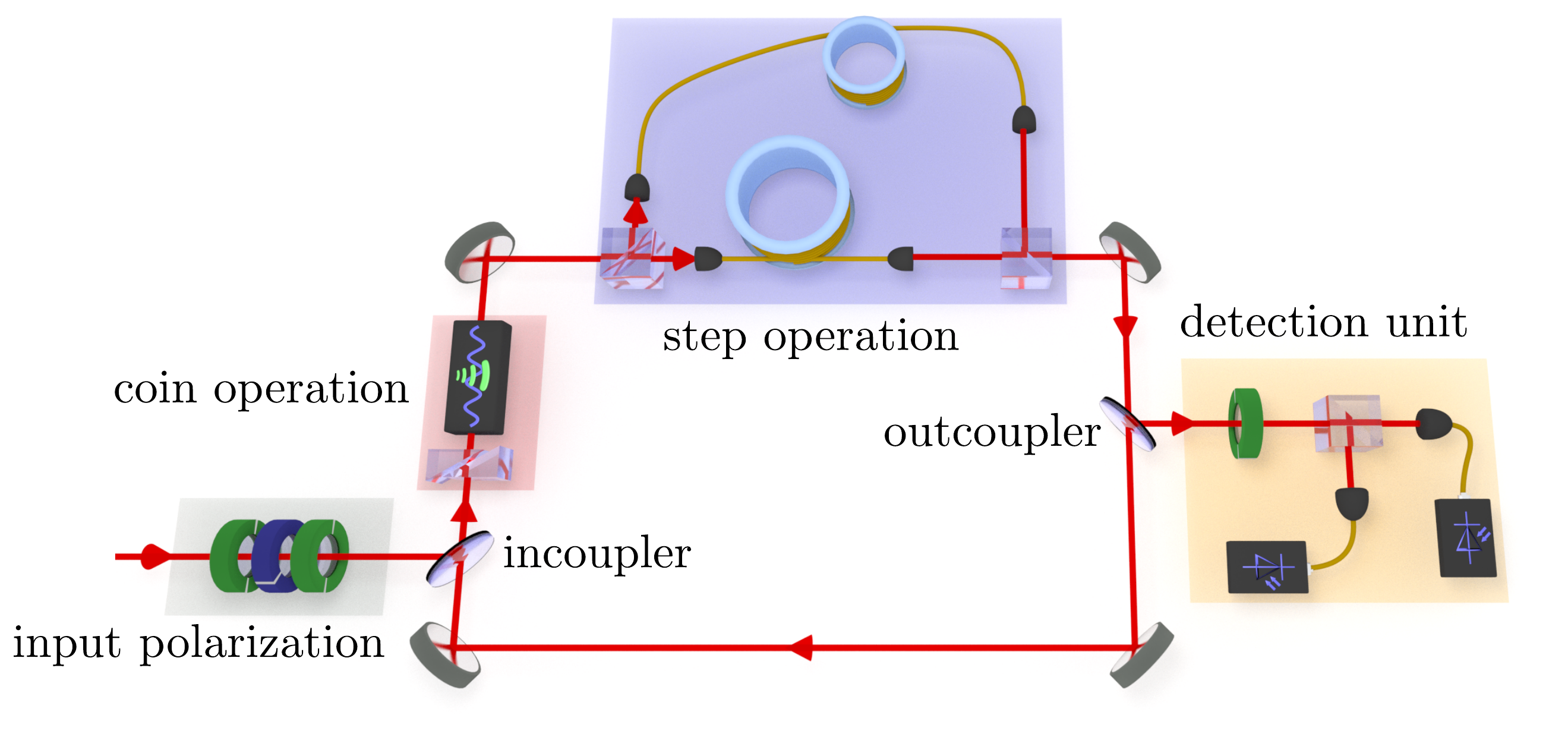}   
 \caption{\label{fig:Exp_Setup} Setup sketch of the quantum walk, for the description see text.
}
\end{figure}

The complete setup is depicted on Fig.~\ref{fig:Exp_Setup}.
A Soleil--Babinet compensator (SBC) and electro-optic modulator (EOM) realise the dynamic coin operation (red area) according to the matrix eq.~(\ref{eq:R_EOM}).
Two standard single-mode fibres of different lengths and two polarizing beam splitters (PBS) implement the step operator (blue area).
A quarter wave plate in front of the detection unit comprising two avalanche photodiodes and a PBS for polarization resolving measurements (orange area) performs a basis transform for the extraction of relative phases, as described at the end of this section.
The three input waveplates can generate an arbitrary polarized input state and compensate for polarization rotations of the incoupler (grey area). 

The transformation realised by the EOM, up to a constant phase factor, corresponds to a Pauli $x$ rotation
\begin{equation}
  \hat{R}(\Phi) = e^{-i\hat{\sigma}_x \Phi} =
  \left(
  \begin{array}{cc}
    \cos \Phi & -i \sin \Phi \\
    -i \sin \Phi & \cos \Phi \\
  \end{array}
  \right),
  \label{eq:R_EOM}
\end{equation}
in the H-V basis \cite{nitsche_quantum_2016}.
The rotation angle $\Phi$ is controlled by applying a voltage $U$.
With a reference voltage $U_0$ preset for each experiment, the EOM can be dynamically switched between three voltages, $U = 0, \pm U_0$, corresponding to rotation angles $\Phi_U = 0, \pm \Phi_{U_0}$.
In combination with a Soleil--Babinet Compensator (SBC), which realises a static polarisation rotation, we are able to implement dynamical switching between the three coin operations $\hat{R}(\Phi_{\mathrm{SBC}} + \Phi_{U})$,
\begin{eqnarray}
  \hat{C}_{\mathrm{U = 0}} &\equiv& \hat{R}(0) \hat{R}(\Phi_{\mathrm{SBC}})= \hat{R}(\Phi_{\mathrm{SBC}}), \label{eq:Coins} \\
  \nonumber \hat{C}_{\mathrm{U = \pm U_0}} &\equiv& \hat{R}(\pm \Phi_{U_0}) \hat{R}(\Phi_{\mathrm{SBC}})= \hat{R}(\Phi_{\mathrm{SBC}} \pm \Phi_{U_0}),
\end{eqnarray}
with two continuous control parameters $\Phi_U$ and $\Phi_{\mathrm{SBC}}$.

A particular spatial coin distribution ${\bm \theta}$ can be achieved by appropriately adjusting the voltage and programming the switching times of the EOM to address the corresponding pulses, indispensable for the realization of the three systems depicted on Fig.~{\ref{fig:leadBulk}.
In order to realise a split-step quantum walk according to eq.~(\ref{eq:splitstepU}) two roundtrips in the fibre loop are required.
Coupling out a small portion of the light probabilistically in each step and measuring it by a pair of avalanche photo diodes (APD) with a large dynamic range allows us to observe the polarization resolved time evolution of the walker. 

To determine the topological invariants according to eq.~(\ref{eq:refl_matrix}) we need to measure not only the absolute values of the reflection amplitudes but also their phases.
Since with the given coin operators the reflection amplitudes $r_j$ are purely imaginary \cite{tarasinski_scattering_2014} which conveys the underlying symmetry of the system, only their signs need to be determined in addition to the intensities.
We have implemented a scheme to discriminate the relative signs of pulses corresponding to neighbouring sites (see Fig.~\ref{fig:relPhases}).
The scheme employs two types of intensity measurements.
In the first, the intensity information of the reflection amplitude is measured (dashed detector positions), while the second type of measurement is carried out after an additional roundtrip and is used to extract the sign information.
In this additional round trip we create interference between successive reflection amplitudes by switching the EOM to a mixing coin $\hat{R}(\alpha)$ set either to $\alpha=\theta_1$ or  $\alpha=\theta_2$  (blue diamonds, $\theta_2$, on Fig.~\ref{fig:relPhases}) instead of the identity coin applied in the lead.
The resulting pulses are sent to a detection unit consisting of a tomography quarter wave plate (QWP) at $45^\circ$, a polarizing beam splitter and two APDs.
The relative sign between the two pulses is extracted from the intensity difference in the two detectors (see appendix~\ref{App:sign-extraction} for the explicit formulas).
To determine the global sign associated to the reflected pulses we have placed a known QWP outside the loop corresponding to a coin operation $\hat{C}_\mathrm{ext}$, thereby splitting off part of the input pulse to be used as an independent reference pulse.
With respect to this reference all measured reflection matrix elements turn real, in accordance with the uniqueness of the invariants $Q_0$ and $Q_{\pi}$ up to a multiplicative factor \cite{tarasinski_scattering_2014}.
\begin{figure}[h!tb]
 \centering  
 \includegraphics[width=0.9\columnwidth]{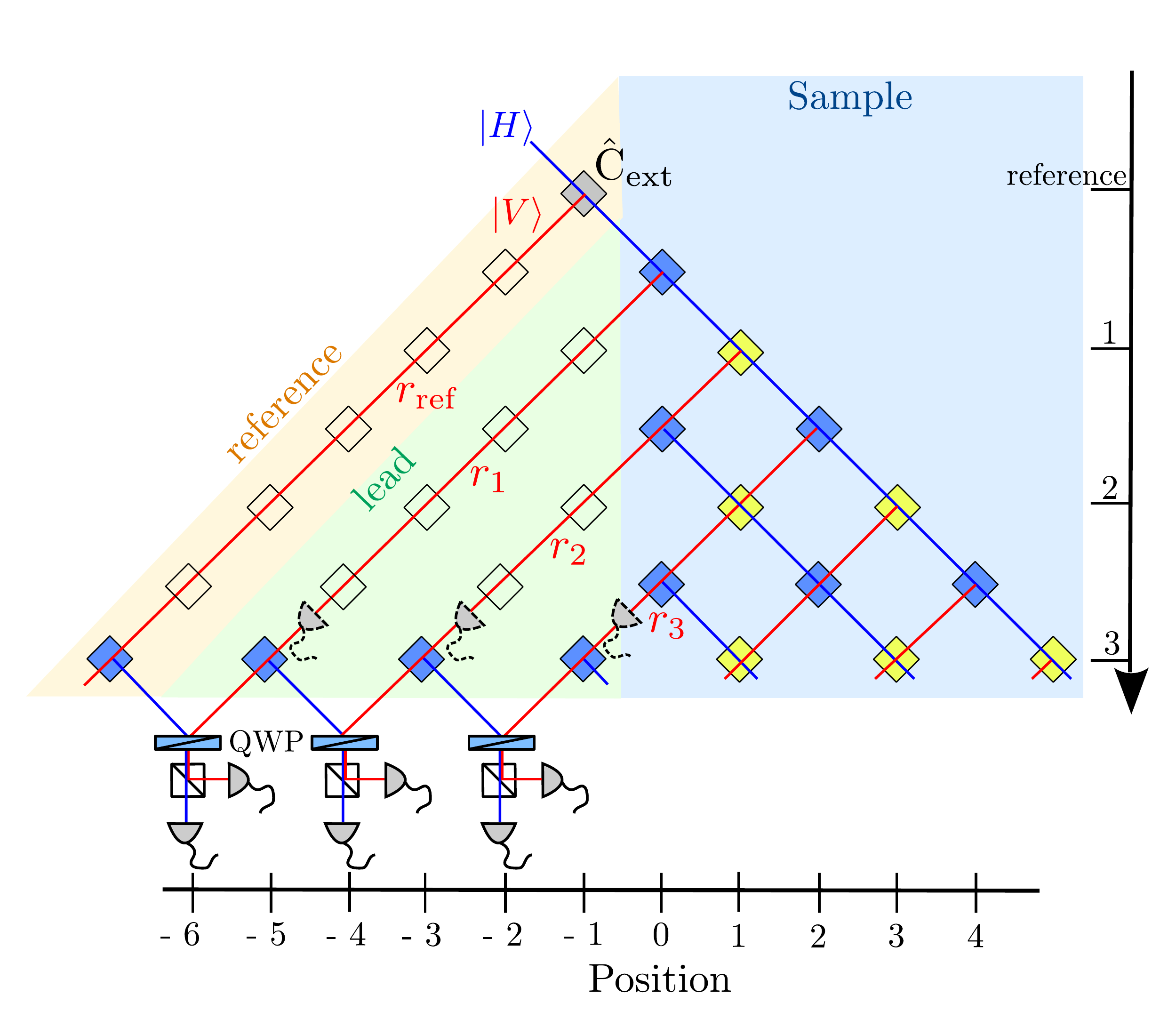}   
 \caption{\label{fig:relPhases} Scheme of the phase extraction for a lead-sample scattering system with external reference signal (here exemplary after 3 steps), showing the relevant measurements. 
The intensity of the reflection amplitudes $|r_j|^2$ is determined using the detectors indicated by dashed lines, while the relative signs of neighbouring positions are extracted one roundtrip later in a tomographic measurement indicated by detectors drawn with solid lines.
The transparent diamonds correspond to the identity coin in the lead region (left half, green shaded) and the blue (dark) and yellow (light) diamonds to $\hat{C}_1$ and $\hat{C}_2$, respectively (sample, right half, blue shaded). 
The possible paths of the two polarizations are marked in red (vertical) and blue (horizontal), respectively. 
The external coin $\hat{C}_{\mathrm{ext}}$ placed outside the loop and used to split off the phase reference is marked by a grey diamond (orange shaded area).}
\end{figure}

\section{Results}
In the first series of measurements, we experimentally analyse the topological invariants for samples without disorder in the arrangement shown on Fig.~\ref{fig:leadBulk}a.
The scenario requires that one of three coin operators is the identity $\openone$.
To implement this operation we match the SBC rotation angle $\Phi_{\mathrm{SBC}}$ to the EOM angle for every preset voltage $U_0$, such that $\Phi_{\mathrm{SBC}} = \Phi_{U_0}$.
This ensures that the negative voltage $-U_0$ yields the Pauli rotation angle $\Phi_{\mathrm{SBC}} - \Phi_{U_0} = 0$.
Consequently, the other two available coin angles then obey $\theta_1=2\theta_2$ with $\theta_2=\Phi_{U_0}$ or $\theta_2=2\theta_1$ with $\theta_1=\Phi_{U_0}$ (green line on Fig.~\ref{fig:parameterPlot}), or $\theta_2=2\theta_1$ with $\theta_1=\Phi_{U_0}$ (turquoise line).
Scanning along each of these two lines is possible by adjusting the EOM voltage $U_0$.
\begin{figure}
 \centering
 \includegraphics[width=\columnwidth]{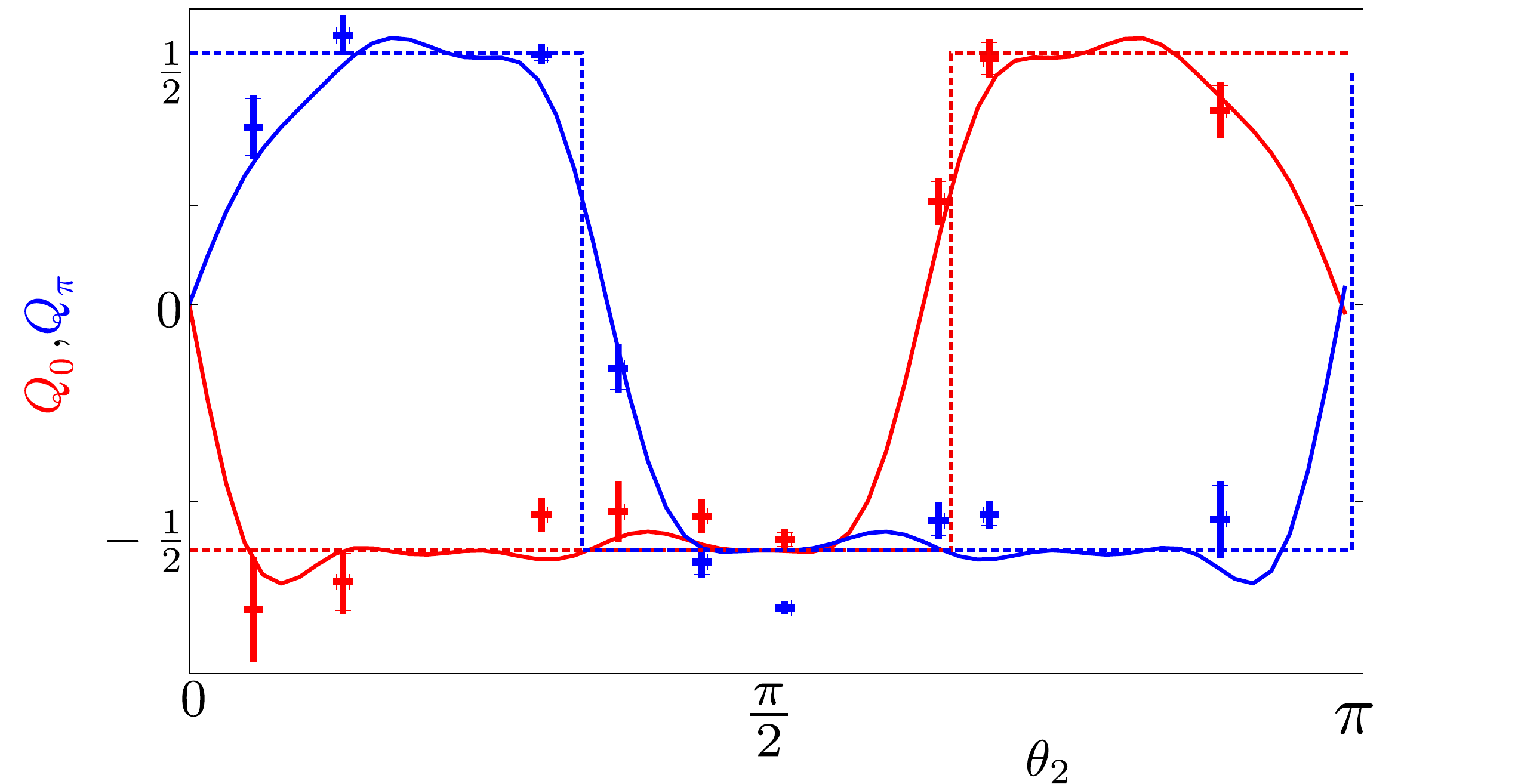}     
 \caption{\label{fig:Exp_Inv} Pair of topological invariants $(Q_0,Q_{\pi})$ (red/light and blue/dark colours, respectively) witnessing topological phase transitions, determined using the scheme on Fig.~\ref{fig:leadBulk}a with coin angles $\theta_1$ and $\theta_2$ scanned along green line of Fig.~\ref{fig:parameterPlot}. The experimental and simulation results for 5 steps of the split-step walk are shown with circles and solid lines, respectively. The expected theoretical bulk topological invariants for infinite systems are marked with dashed lines. Error bars are obtained by a Monte Carlo simulation (see main text). For results from scanning along the turquoise line consult appendix~\ref{App:system2}.}
\end{figure}
The intensity amplitudes are measured after roundtrip 9 and the relative signs of the reflection amplitudes are extracted in roundtrip 10 as shown in Fig.~\ref{fig:relPhases} without the reference pulse, corresponding to performing 5 steps of the split-step walk.
The reflection amplitudes $r_j$ are thus reconstructed up to a global sign, allowing us to determine the approximate absolute value and relative sign of the topological invariants $Q_0$ and $Q_{\pi}$ using eq.~(\ref{eq:refl_matrix}).
This information is sufficient to identify all topological phase transitions occurring along the green and turquoise lines of Fig.~\ref{fig:parameterPlot}.
The measurement results for the green line scan are shown on Fig.~\ref{fig:Exp_Inv}, along with the simulations for 5 steps (solid lines), and the expected theoretical values for infinite samples (dashed lines) of the bulk topological invariants (the global phases of the measured $Q_0$ and $Q_{\pi}$ are chosen to allow for easy comparison).
The positions of the observed phase transitions agree well with the theory, and the simulations show that the deviation is dominantly due to finite size effects, which smear out the sharp transitions from one phase to the other.
The errorbars are obtained via a Monte-Carlo simulation taking the experimental error sources like losses and angle misalignment of the EOM crystal and the SBC into account, see appendix~\ref{App:errorbars}.

The study of disordered samples is motivated by the predicted robustness of the topological invariants against certain types of disorder, bringing us to our second series of experiments.
To test this property we realize large ensembles of disordered samples experimentally, making use of the easy reprogrammability of the EOM.
As described above we are restricted to three coin angles, thus we consider samples which have $\hat{C}_1 = \hat{\openone}$ while $\hat{C}_2$ is spatially varied according to Fig.~\ref{fig:leadBulk}(b).
As a practical simplification the identity operations $\hat{C}_1=\openone$ (both on the lead and the sample side) are taken into account implicitly, allowing us to realize 11 steps of the split-step walk in 11 round trips, since every odd roundtrip would only comprise identity operations and thus can be skipped.
The third available coin operation is still at hand and used to introduce disorder as follows.
For each $\theta_x$ (in the sample) we randomly choose from $\theta_{\mathrm{A}}=\theta$ and $\theta_{\mathrm{B}} = 2\theta$ with the respective probabilities $1-p$ and $p$.
Therefore, $p=0.5$ corresponds to maximum disorder, while $p=0$ and $p=1$ to the disorder-free samples $\hat{C}_2=\openone_x\otimes \hat{R}(\theta_{\mathrm{A}})$  and $\hat{C}_2=\openone_x\otimes \hat{R}(\theta_{\mathrm{B}})$, respectively.
For this lead--sample system the formula eq.~(\ref{eq:refl_matrix}) yields identical results for $r(0)$ and $r(\pi)$ as every second summand is zero, and therefore we will analyse only $r(0)$ in the following.
\begin{figure}
 \centering
 \includegraphics[width=\columnwidth]{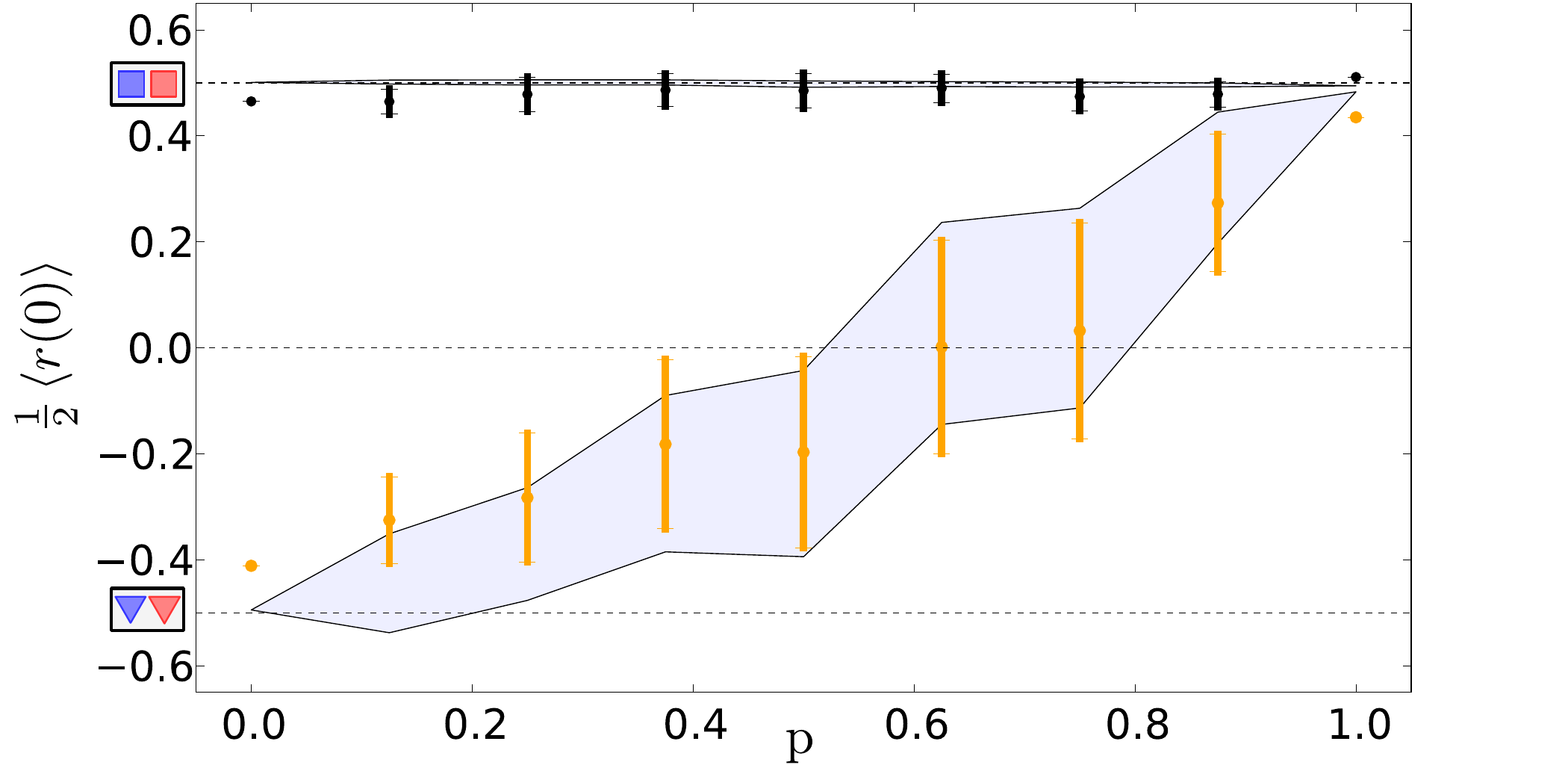}   
 \caption{\label{fig:Exp_Noise} The (scaled) average reflection matrix element $\frac12\left<r(0)\right>$ for disordered samples after 11 steps; to be compared to the respective bulk topological invariants, $Q_0 = \pm \frac12$ (indicated by dashed lines). Black symbols: first case study, represented by squares in Fig.~\ref{fig:parameterPlot} at $\theta_{\mathrm{A}} = 1.68\pi$ and $\theta_{\mathrm{B}} = 1.36\pi$. Orange/light symbols: second case study, represented by triangles in Fig.~\ref{fig:parameterPlot} at $\theta_{\mathrm{A}} = 0.63\pi$ and $\theta_{\mathrm{B}} = 1.26\pi$. The grey shadings indicate the range of the standard deviations of the simulations. Error bars are given by the standard deviation of the 50 measured patterns for each probability $p$. Dashed line at $0$ is a guide to the eye.
   }
\end{figure}

In Fig.~\ref{fig:Exp_Noise} we present results from two case studies of disordered systems: For each value of $p$ (except the unperturbed systems with $p = 0$ and $p= 1$) we measured 50 randomly generated coin distributions and calculated the ensemble average reflection matrix $\left<r(0)\right>$ after 11 steps (see eq.~(\ref{eq:refl_matrix})), which we compare to the topological invariant $Q_0$.
Here, we created a reference signal prior to the first step to extract the global phases of the reflection amplitudes, see appendix~\ref{App:sign-extraction}.

In the first case study (black symbols in Fig.~\ref{fig:Exp_Noise}) we used coins with angles $\theta_{\mathrm{A}} = 1.68\pi$ (red square in Fig.~\ref{fig:parameterPlot}) and $\theta_{\mathrm{B}} = 1.36\pi$ (blue square), both of which belong to the same topological phase with bulk topological invariants $(Q_0,Q_{\pi})=(\frac12, \frac12)$.
As expected, the observed value $\frac{1}{2}\left<r(0)\right>\approx \frac{1}{2}$ is approximately independent of the disorder parameter $p$, confirming that all disordered systems belong to the same topological phase $Q_0=\frac12$: The value $\left<r(0)\right>$ is invariant with respect to this kind of disorder.
The statistical fluctuations (black error bars) of the 50 ensembles are very small and caused by finite size effects and inevitable experimental inaccuracies, fully in agreement with the simulation (grey shadow).

The second case study (orange symbols in Fig.~\ref{fig:Exp_Noise}) contains coins with $\theta_{\mathrm{A}} = 0.63\pi$ and $\theta_{\mathrm{B}} = 1.26\pi$ (resp.\ red and blue triangles on Fig.~\ref{fig:parameterPlot}), belonging to the different topological phases $(-\frac12, -\frac12)$ and $(\frac12, \frac12)$, respectively.
Using the methods developed in \cite{rakovszky_localization_2015} for the determination of Lypunov exponents and localisation length for systems with binary disorder theoretical calculations for the \emph{infinite} system indicate a sharp transition from $-\frac12$ to $+\frac12$ at $p_{\mathrm{crit}} = 0.625$ as we increase $p$ from 0 to 1.
However, the sum of eq.~(\ref{eq:refl_matrix}) with a finite step number of $t = 11$ steps has not converged yet, thus for our observation time this effect is smeared out.
We observe a smooth transition of $Q_0$ from $-\frac12$ to $\frac12$ and large statistical fluctuations for the 50 ensembles (orange error bars), which is both fully reproduced in the finite-size simulations (grey shaded area).
The great agreement of the experimental data with the numerical simulations of the system, given the same finite observation time proves that only the slow convergence of eq.~(\ref{eq:refl_matrix}) is the cause for the smoothing and no experimental failures suddenly occurred.
The difference between the two curves is obvious: While for the first disorder case (black curve) the value of $\left<r(0)\right>$ is constant no matter how strong the disorder is, the second curve (orange) shows a clear dependence on the disorder parameter $p$ and $\left<r(0)\right>$ is not constant.

\begin{figure}
 \centering
\includegraphics[width=\columnwidth]{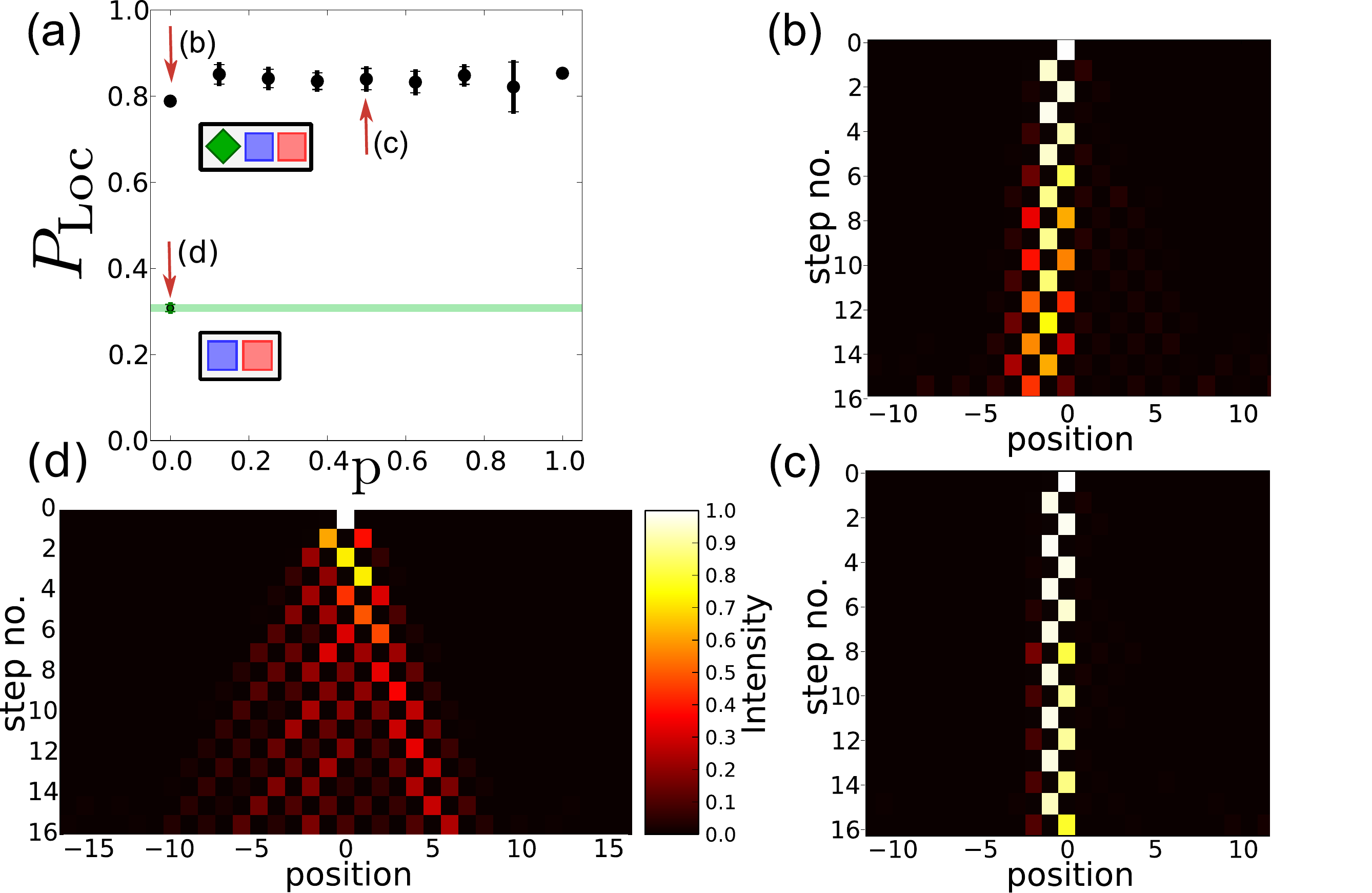}	
 \caption{\label{fig:Exp_Loc} (a) Black symbols: Localisation as a probability of finding the walker in the central bins versus disorder strength $p$ for the bulk--bulk interface (Fig.~\ref{fig:leadBulk}~(c)) with $\theta_{\mathrm{A}} = 1.68\pi$, $\theta_{\mathrm{B}} = 1.36\pi$ and $\theta_L = 0.52\pi$. Error bars are given by the standard deviation of the 50 different ensembles for each probability $p$. 
Green symbol: reference system ($\theta_A = 1.36\pi$ and $\theta_L = 1.68\pi$) without localisation extended by a green shaded area indicating the reference value including systematic errors obtained by a Monte-Carlo simulation to allow for comparison, see appendix~\ref{App:errorbars} for details. Panels (b),(c),(d) show three wavefunctions for the configurations indicated by the arrows (bright colours denote high, and dark colours low intensities).}
\end{figure}

In our last series of experiments, we investigate edge states occurring on the boundary of two samples.
In the experimental scenario, depicted on Fig.~\ref{fig:leadBulk}(c), a homogeneous sample with non-trivial coin is implemented on the left, while the right sample is disordered.
The samples on the right are identical to those which have been studied in the first case study above, i.e.\ the coin distribution consisting of 
the angles $\theta_{\mathrm{A}} = \theta = 1.68\pi$ and $\theta_{\mathrm{B}} = 2\theta = 1.36\pi$ chosen randomly.
The control parameter $p$ allows us to tune the extent of disorder, and study the robustness of the edge states against it.
The left sample comprises only one coin angle $\theta_L = 3\theta/2 = 0.52\pi$ (squares in Fig.~\ref{fig:parameterPlot}).
In a separate scattering experiment we have obtained the value of $\frac12 r(0) = -0.581 \pm 0.03$ for the scaled reflection matrix, revealing the topological invariant $Q_0=-\frac12$ (up to systematic errors) for the left sample.
Consequently, the full system comprises the interface of two topologically distinct samples, independent of the disorder parameter $p$, is therefore expected to support edge states on the boundary.
In 1-dimensional disordered systems, especially when only the first steps are available, it is impossible to distinguish between localised states of topological origin and those caused by Anderson localisation \cite{anderson_absence_1958} without determining the explicit eigenvalues of the states \cite{obuse_topological_2011, rakovszky_localization_2015}. 
Thus in the following we present a purely experimental comparison of such localised states of disorder-free samples with topological phase boundary, where no Anderson localisation occurs ($p=0$ and $p=1$), of samples with disorder and phase boundary ($p\ne 0,1$), as well as a reference system with neither disorder nor phase boundary.
A more thorough study of the localisation properties will be given in \cite{gabris_2017}.

We quantify the degree of localisation by $P_{\mathrm{Loc}} = \sum_{i = -3}^3 P_i$, the probability of finding the walker in the central positions within the interval $[-3,3]$.
After 13 steps, we compare the degree of localization to the reference system exhibiting no localization, consisting of two samples in the same topological phase with respective angles $\theta_L = 1.68\pi$ and $\theta_R = 1.36\pi$ (red and blue squares in Fig.~\ref{fig:parameterPlot}).
The experimental results for the degree of localization are presented on Fig.~\ref{fig:Exp_Loc}(a).
We clearly see that the localisation strength $P_{\mathrm{Loc}}$ involving a phase boundary (black symbols) is almost three times higher than for the reference system (green symbol), even when systematic errors via a Monte-Carlo simulation are included (green shaded area).
The localisation strength remains roughly constant while increasing the disorder from the ordered system at $p = 0$ to the maximum disorder at $p = 0.5$ back to the second ordered system at $p = 1$.
The statistical fluctuations of the 50 configurations per probability $p$ (black error bars) are small compared to the mean values, proving that each single configuration contains a localised state.

The panels (b),(c),(d) of Fig.~\ref{fig:Exp_Loc} show the three selected wavefunctions in an intensity plot.
The presented data sets are concatenated from a low power measurement for the first 6 steps and a high power measurement for the higher steps to avoid detector saturation effects.
The two wavefunctions (b) and (c) for the localised system exhibit the expected peak of intensities at the boundary between the two samples at $x = 0$, and although belonging to the two extremal disorder cases $p=0$ and $p=0.5$, their shapes do not differ significantly.
On the other hand the wavefunction (d) of the non-localised reference system features the usual double-lobe distribution, each lobe ballistically travelling at the speed determined by the supporting sample.

\section{Conclusion}
Using the time-multiplexing DTQW setup we experimentally studied features of topological insulators, in particular we focused on direct measurements of the topological invariants and edge states in ideal and disordered samples.
The easy reconfigurability and programmability of the setup enabled us to precisely control and tune the disorder strength in ensemble averages over in total several hundred configurations.
The full dynamic control of the system parameters combined with the loop extension already demonstrated for 2-dimensional quantum walks \cite{schreiber_2d_2012} may also provide a promising approach for the realisation of complex dynamics such as the excitation and evolution of topologically protected edge states in higher dimensions.


\begin{acknowledgments}
This work has received funding from the European Union’s Horizon 2020 research and innovation programme under the QUCHIP project GA no.\ 641039.
S.~B., T.~N., F.~E., L.~L.\ and C.~S.\ acknowledge funding by the DFG (Deutsche Forschungsgemeinschaft) via the Gottfried Wilhelm Leibniz-Preis.
I.~J.\ has been partially supported by the Czech Science foundation (GA{\v C}R) project number\ 16-09824S, A.~G.\ and I.~J.\ from No.\ GA{\v C}R 13-33906S, RVO 68407700, and A.~G.\ from T\'AMOP-4.2.4.A/2-11/1-2012-0001 ``National Excellence Program'' of Hungary.
\end{acknowledgments}


\begin{appendix}

\section{Topological Invariants for a second System}
\label{App:system2}
Just by reprogramming the switching times of the EOM we can change the order of $\theta_1$ and $\theta_2$.
The measured topological invariants are plotted in Fig.~\ref{fig:pattern01}.
This pattern corresponds to scanning along the turquoise line in the parameter plot (Fig.~2, main text).
Here the positions of the phase transitions between same sign and opposite sign of $Q_0$ and $Q_{\pi}$ take place at different angles compared to Fig.~3 in the main text.
We find the same sign for $\theta_1 = 0.73\pi$, opposite sign for $\theta_1 = 1.12\pi$ and a transition region for $\theta_1 = 0.90\pi$ in accordance with the finite-step simulation.
\begin{figure}[h]
\centering
 \includegraphics[width=0.95\columnwidth]{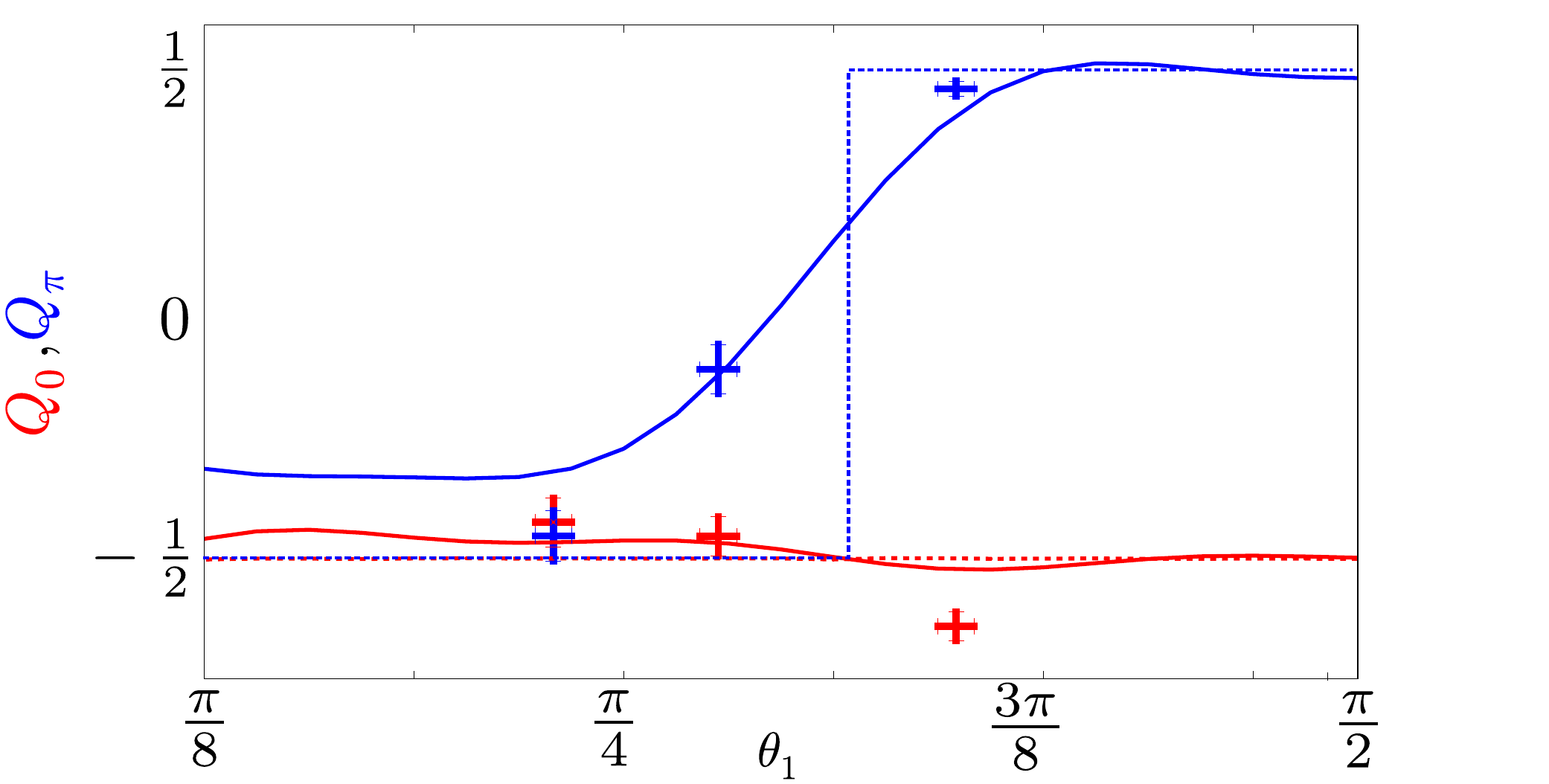}  
 \caption{\label{fig:pattern01} Topological invariants $Q_0$ (red/light symbols) and $Q_{\pi}$ (blue/dark symbols), simulations for 5 steps of the split-step walk (solid lines), and theoretical values of the bulk topological invariants (dashed lines) for a scan along the turquoise line in Fig.~2 in the main text. Error bars are obtained by a Monte Carlo simulation (see section \ref{App:errorbars} for details).} 
\end{figure}

\section{Sign Extraction}
\label{App:sign-extraction}
In order to extract the topological invariants we need to measure the reflection amplitudes, i.e.\ the intensities as well as the phases according to eq.~(6) in the main text.
For the given coin operators (see eq.~(\ref{eq:Coins})) the reflection amplitudes are purely imaginary \cite{tarasinski_scattering_2014} therefore only their sign needs to be determined.
In the experiment, phase measurements have to be carried out separately as the APDs only give us the intensity information.
Therefore, we have included an additional transformation to create interference between pulses belonging to neighbouring positions to determine their relative signs.
For example, to obtain full amplitude information for the 3rd step, we use EOM settings equivalent to the beam-splitter cascade on Fig.~\ref{fig:relPhases}.
Note that the figure shows two intensity measurements: the intensities are extracted one roundtrip before the last one, while the sign information is determined in a second measurement carried out in the final roundtrip.
The interference between successive reflection amplitudes is realised by switching the EOM to a mixing coin $\hat{R}(\alpha)$ instead of the identity in the lead region (blue coins).
Again, the choice of $\alpha$ is restricted by the EOM capabilities to the angles corresponding to $\hat{C}_1$ or $\hat{C}_2$ (yellow and blue diamonds).
After two neighbouring pulses interfered in succession of the corresponding coin operation we couple them out (as in every step) and route them to the detection unit.
The pulses pass then a tomography QWP at $45^\circ$, and are afterwards measured in a polarization resolved way.
The relative sign between the two pulses can be extracted from the intensity difference in the H- and V-detectors as shown in the following calculus:
The complex wave function at e.g.\ position $x=-4$ is given by
\begin{eqnarray} \nonumber
|\Psi\rangle_{-4} &=& \hat{C}_{\mathrm{QWP}}\left[\hat{P}_H\hat{R}(\alpha)
\begin{pmatrix}
0 \\
r_1 \\
\end{pmatrix}
 + \hat{P}_V\hat{R}(\alpha)
\begin{pmatrix}
0 \\
r_2 \\
\end{pmatrix}\right]
\\ \nonumber
&=& \frac{1}{\sqrt{2}}
\begin{pmatrix}
1 && i \\
i && 1 \\
\end{pmatrix}
\Big[
\hat{P}_H
\begin{pmatrix}
\cos\alpha && -i\sin\alpha \\
-i\sin\alpha && \cos\alpha \\
\end{pmatrix}
\begin{pmatrix}
0 \\
r_1 \\
\end{pmatrix} 
\\ \nonumber
 &&~~~~~~~~~~~~~~~~~~~+ 
\hat{P}_V
\begin{pmatrix}
\cos\alpha && -i\sin\alpha \\
-i\sin\alpha && \cos\alpha \\
\end{pmatrix}
\begin{pmatrix}
0 \\
r_2 \\
\end{pmatrix} \Big]
\\ \nonumber
&=& 
\frac{1}{\sqrt{2}} 
\begin{pmatrix}
-ir_1\sin\alpha +ir_2\cos\alpha \\
r_1\sin\alpha+r_2\cos\alpha \\
\end{pmatrix}
\label{eq:signs}
\end{eqnarray}
where 
$\hat{P}_H = \begin{pmatrix}
1 && 0 \\
0 && 0 \\
\end{pmatrix}$ 
and 
$\hat{P}_V = \begin{pmatrix}
0 && 0 \\
0 && 1 \\
\end{pmatrix}$ are the projectors on the horizontal and vertical basis element, respectively, and $r_j$ the (vertically polarized) reflection amplitudes.
This means that the intensity measured in the H-(V-) detector behind the PBS reads
\begin{eqnarray} \nonumber
I_{H/V} = \frac{1}{2}\left( r_1^2\sin^2\alpha \mp 2 r_1 r_2 \sin\alpha \cos\alpha + r_2^2\cos^2\alpha \right).
\end{eqnarray}
It follows, that by knowing the intensity difference,
\begin{eqnarray}
\Delta I = I_H - I_V = -2 r_1 r_2 \sin(\alpha)\cos(\alpha),
\label{eq:DeltaI}
\end{eqnarray}
of the two detectors, and the EOM angle $\alpha$ one can determine the relative signs of $r_1$ and $r_2$.
In particular for $\alpha \in [0,\pi/4]$, if the intensity in the H-detector is smaller than in the V-detector then $r_1$ and $r_2$ have the same sign, while the intensity is larger in the H- than in V-detector we have $\mathrm{sign}(r_1)\ne \mathrm{sign}(r_2)$.
In this way we can sequentially determine every sign relation between every successive reflection amplitude $r_j$ and $r_{j+1}$, enabling us to reconstruct the sign of every amplitude relative to the first one.

In order to determine the global sign of all amplitudes, it is necessary to use a reference pulse.
We have created the reference pulse by putting a known wave plate aligned at well-defined angle, realizing the transformation $\hat{C}_{\mathrm{ext}}$ (grey diamond), before the input is coupled into the loop.
Thus, the reference pulse with known phase from the input state (orange shaded region) is split off in the first roundtrip.
At the price of this additional roundtrip we are able to compare the first reflection amplitude $r_1$ with the reference pulse and by this means obtain the global sign.

Since only sign changes between neighbours can be extracted, the described method is very sensitive to errors: if one sign is extracted the wrong way all following signs will also be wrong.
This issue becomes most serious for angles $\alpha$ close to multiples of $\pi/2$, because then the intensity difference eq.~(\ref{eq:DeltaI}) goes to zero.
The same is true if one or both of the amplitudes $r_j$ become very small.
For the presented results on the stability properties of the topological invariants in Fig.~\ref{fig:Exp_Inv} in the main text, we successfully performed the sign extraction in step 5 (which means after 10 roundtrips) which proves the high accuracy of the experimental setup.

\section{Errorbars}
\label{App:errorbars}
We have identified four sources of systematic errors in our experimental setup, compare also \cite{elster_quantum_2015}: first, the detector and power dependent detection efficiencies, which were determined in a separate measurement; 
second, the different losses experienced in different paths due to dissimilar coupling efficiencies and path geometries, which are estimated in an independent measurement with an accuracy of $\pm 3\,\%$; 
third, the exact angle of the (switched) EOM which can only be determined up to $\pm 1^\circ$;
fourth, the angle of the SBC can be set only with a precision of $1^\circ$.

For the determination of the parameters of the other three errors we resorted to a numerical model.
In a Monte Carlo simulation we randomly chose 1000 sets from the parameters within the identified ranges. 
The set yielding the best reproduction of the experimental data (we calculated the distance between simulation and experiment for the first 7 roundtrips) was chosen for the realistic model.
The mean deviation of the statistics produced by the Monte Carlo simulation from the realistic model determines the size of the presented errorbars.

\section{Split step scheme}
\label{App:splitstepscheme}
For the investigation of topological effects in DTQW originally a split-step scheme consisting of two coin operations in between the asymmetric step operators $\hat{S}_\pm$ was proposed \cite{kitagawa_exploring_2010, tarasinski_scattering_2014}.
It is fully equivalent to two steps with the symmetric step operator $\hat{S}$ and alternating coins when relabelling the positions in every second step, as demonstrated in Fig.~\ref{fig:splitstep}.

\begin{figure}[h]
 \centering
 \includegraphics[width=0.95\columnwidth]{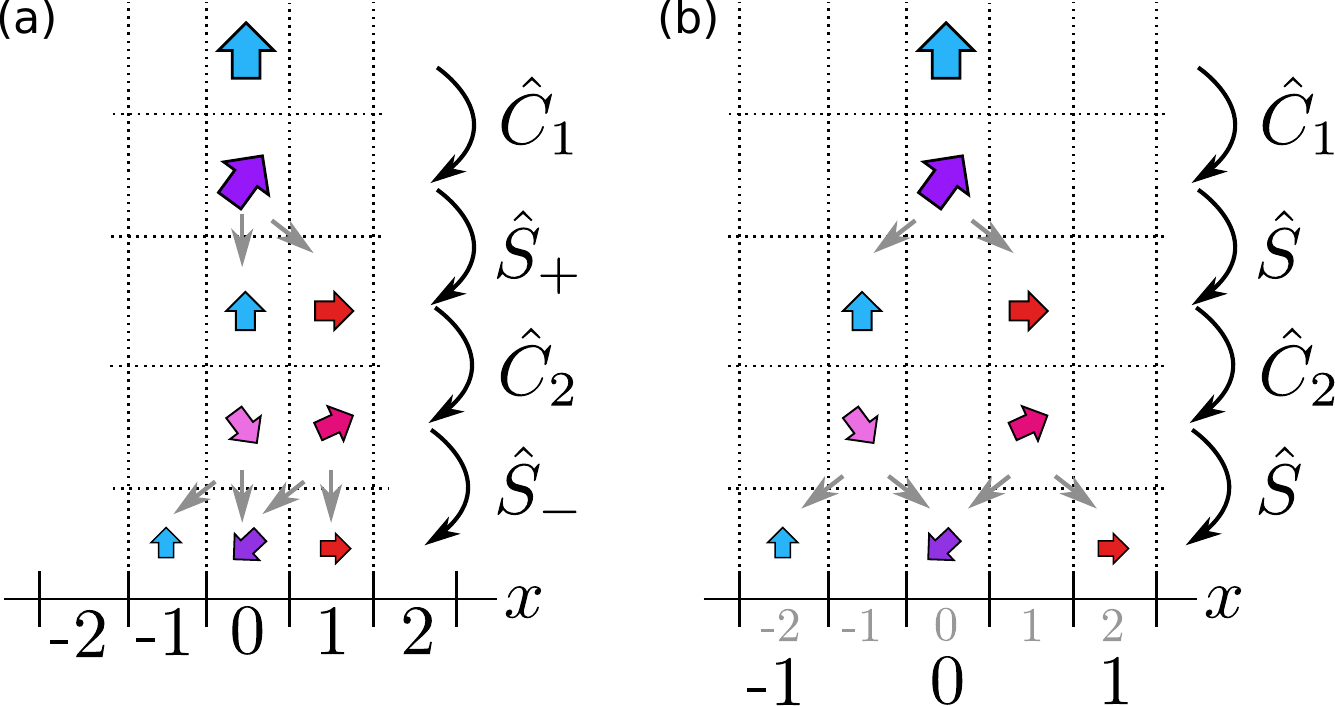}	
 \caption{\label{fig:splitstep} Exemplary evolution of a quantum walker's wavefunction (coloured arrows) (a) Originally proposed split-step scheme according to $\hat{U} =  \hat{S}_- \hat{C}_2 \hat{S}_+ \hat{C}_1$; (b) Experimental realisation of the split step scheme using a double step protocol governed by $\tilde{U} = \hat{S} \hat{C}_2 \hat{S} \hat{C}_1$ with subsequent position relabelling (original labels: grey; after relabelling: black); for simplicity, we chose a vertically polarized input state (blue arrow in the first line)}
\end{figure}
\end{appendix}

%

\end{document}